\documentclass[prl,twocolumn,superscriptaddress,showpacs,amscd,amsmath,amssymb,verbatim]{revtex4-1}

\usepackage{graphicx}
\usepackage{wrapfig}
\usepackage{epsfig}
\graphicspath{ {./images/} }
\usepackage{dcolumn}
\usepackage{bm}
\usepackage{amsmath}
\usepackage{gensymb}
\usepackage{epstopdf}
\usepackage{cuted}
\usepackage{color}
\usepackage{mathtools}
\usepackage{upgreek}
\usepackage{natbib}
\usepackage{hyperref}
\hypersetup{colorlinks=true, linkcolor=blue, citecolor=blue, urlcolor=blue}
\usepackage[mathlines]{lineno}
\usepackage{comment}

\usepackage[normalem]{ulem}
\usepackage[usenames, dvipsnames]{xcolor}

\newcommand{\beginsupplement}{
        \setcounter{table}{0}
        \renewcommand{\thetable}{S\arabic{table}}
        \setcounter{figure}{0}
        \renewcommand{\thefigure}{S\arabic{figure}}
     }
     
\begin{document}


\title{
 Field-induced double spin spiral 
 in a frustrated chiral magnet
}

\author{M. Ramakrishnan}
\email{mahesh.ramakrishnan@hotmail.com}
\affiliation{Swiss Light Source, Paul Scherrer Institut, 5232 Villigen PSI, Switzerland}
\author{E. Constable}
\email{evan.constable@tuwien.ac.at}
\affiliation{Institute of Solid State Physics, Vienna University of Technology, 1040 Vienna, Austria}
\affiliation{Univ. Grenoble Alpes, CNRS, Institut N\'eel, F-38042 Grenoble, France}
\author{A. Cano}
\email{andres.cano@cnrs.fr}
\affiliation{Univ. Grenoble Alpes, CNRS, Institut N\'eel, F-38042 Grenoble, France}
\affiliation{Department of Materials, ETH Zurich, Vladimir-Prelog-Weg 4, 8093 Zurich, Switzerland.}
\author{M. Mostovoy}
\affiliation{Zernike Institute for Advanced Materials, University of Groningen, Nijenborgh 4, Groningen 9747 AG, The Netherlands}
\author{J. S. White}
\affiliation{Laboratory for Neutron Scattering, Paul Scherrer Institut, 5232 Villigen PSI, Switzerland}
\author{N. Gurung}
\affiliation{Department of Materials, ETH Zurich, Vladimir-Prelog-Weg 4, 8093 Zurich, Switzerland.}
\affiliation{Laboratory for Multiscale Materials Experiments, Paul Scherrer Institut, 5232 Villigen PSI, Switzerland}
\author{E. Schierle}
\affiliation{Helmholtz-Zentrum Berlin f\"{u}r Materialien und Energie, Wilhelm-Conrad-R\"{o}ntgen-Campus BESSY II, Albert-Einstein-Strasse 15, 12489 Berlin, Germany.}
\author{S. de Brion}
\affiliation{Univ. Grenoble Alpes, CNRS, Institut N\'eel, F-38042 Grenoble, France}
\author{C. V. Colin}
\affiliation{Univ. Grenoble Alpes, CNRS, Institut N\'eel, F-38042 Grenoble, France}
\author{F. Gay}
\affiliation{Univ. Grenoble Alpes, CNRS, Institut N\'eel, F-38042 Grenoble, France}
\author{P. Lejay}
\affiliation{Univ. Grenoble Alpes, CNRS, Institut N\'eel, F-38042 Grenoble, France}
\author{E. Ressouche}
\affiliation{Universit\'e Grenoble Alpes, CEA, INAC, MEM, F-38000 Grenoble, France}
\author{E. Weschke}
\affiliation{Helmholtz-Zentrum Berlin f\"{u}r Materialien und Energie, Wilhelm-Conrad-R\"{o}ntgen-Campus BESSY II, Albert-Einstein-Strasse 15, 12489 Berlin, Germany.}
\author{V. Scagnoli}
\affiliation{Department of Materials, ETH Zurich, Vladimir-Prelog-Weg 4, 8093 Zurich, Switzerland.}
\affiliation{Laboratory for Multiscale Materials Experiments, Paul Scherrer Institut, 5232 Villigen PSI, Switzerland}
\author{R. Ballou}
\affiliation{Univ. Grenoble Alpes, CNRS, Institut N\'eel, F-38042 Grenoble, France}
\author{V. Simonet}
\affiliation{Univ. Grenoble Alpes, CNRS, Institut N\'eel, F-38042 Grenoble, France}
\author{U. Staub}
\email{urs.staub@psi.ch}
\affiliation{Swiss Light Source, Paul Scherrer Institut, 5232 Villigen PSI, Switzerland}

\date{\today}

\begin{abstract}

We report the direct observation of a magnetic-field induced long-wavelength spin spiral modulation in the chiral compound Ba$_3$TaFe$_3$Si$_2$O$_{14}$. This 
new spin texture emerges out of a chiral helical ground state, and is hallmarked by the onset of a unique contribution to the bulk electric polarization, the sign of which depends on the crystal chirality. The periodicity of the field induced modulation, several hundreds of nm depending on the field value, is comparable to the length scales of mesoscopic topological defects such as skyrmions, merons and solitons. The phase transition and observed threshold behavior are consistent with a phenomenology based on the allowed Lifshitz invariants for the chiral symmetry of langasite, which intriguingly contain all the ingredients for the possible realization of topologically stable antiferromagnetic skyrmions. 
 

\end{abstract}

\maketitle
 
Non-centrosymmetric magnets are 
outstandingly rich systems for the 
emergence of various types of modulated orders \cite{izyumov84} and topological objects such as skyrmions \cite{fert17}. 
This is related to the presence of  
the so-called Lifshitz invariants, as was originally demonstrated for nominal ferromagnets \cite{Dzyaloshinskii1964,Bak80,bogdanov1989a}. Similar considerations apply to nominally collinear antiferromagnets in the absence of inversion symmetry \cite{bogdanov1989,bogdanov1998,bogdanov2002}. 
Thus, the popular multiferroic material BiFeO$_3$ 
displays a cycloidal long-wavelength modulation of its $G$-type antiferromagnetic order \cite{sosnowska16}, and the theoretically predicted topological defects for collinear antiferromagnets of this class \cite{bogdanov1989} have recently been reported in Ref.~\cite{Seki12}. Furthermore, the challenge of generalizing these concepts to non-collinear antiferromagnets motivates an increasing number of theoretical predictions \cite{cabra15,cabra17} that remain to be experimentally tested. 

\begin{figure}[b!]
\includegraphics[width=0.37\textwidth]{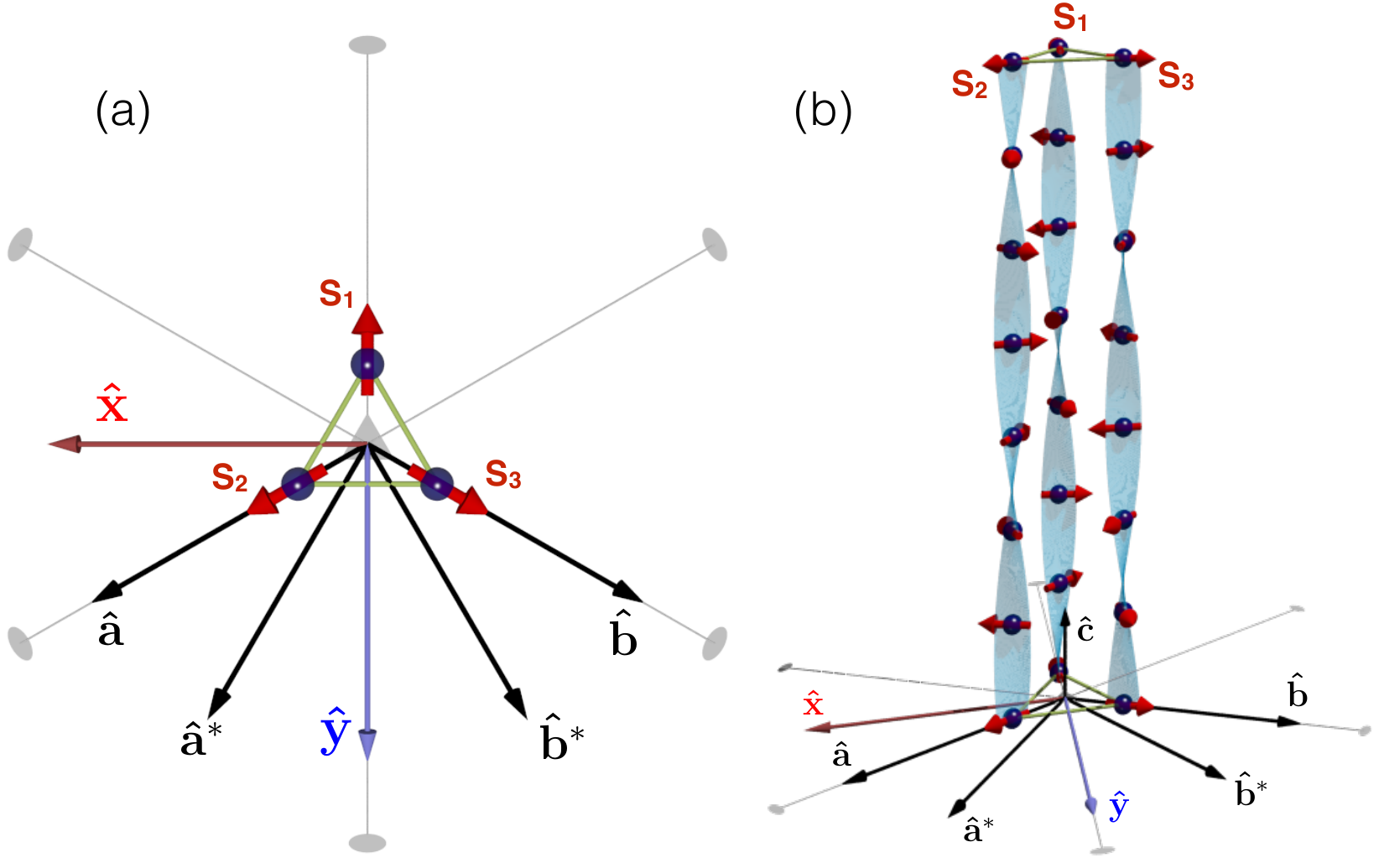}
\caption{Illustration of the 120$^{\circ}$ Fe$^{3+}$ spin order (a) and its helical modulation along the $c$ axis (b) emerging in the Ba$_3M$Fe$_3$Si$_2$O$_{14}$ series. The propagation vector is approximated to (0, 0, 1/7). The different crystallographic axes, as well as the 3-fold and 2-fold rotation axes of the corresponding $P321$ space group are indicated in (a).}
\label{fig1} 
\end{figure}

To what extent these spin textures and their potential chiral character are a source of magnetoelectric couplings and multiferroicity is a pertinent and evolving question. 
In this context, the langasite series with chemical formula Ba$_3M$Fe$_3$Si$_2$O$_{14}$ ($M$ = Sb, Nb, Ta) offers a very promising playground. These multichiral systems crystallize in a chiral structure with $P321$ space-group symmetry, where the Fe$^{3+}$ spins form a triangular lattice in the $ab$ plane \cite{Marty2008,Loire2011,Zorko2011,Scagnoli2013}. 
These spins display an unconventional magnetic order that is essentially determined by frustrated exchange interactions and is peculiarly linked to the chirality of the lattice \cite{Marty2008,Zorko2011,Chaix2016}. 
Specifically, the antiferromagnetic interactions within the Fe$^{3+}$ spin triangles lead to a non-collinear 120$^{\circ}$ order that is helically modulated along the $c$ direction with a periodicity $\sim 7c$ due to competing out-of-plane exchange couplings (see Fig.~\ref{fig1}). The chirality of the 120$^\circ$ order is determined by additional magnetocrystalline single-ion anisotropy and Dzyaloshinskii-Moriya (DM) interactions, which then tie the chirality of the helical modulation to the chirality of the lattice \cite{Scagnoli2013,Chaix2016}. This peculiar magnetic order enables static and dynamical magnetoelectric effects \cite{Marty2010,Chaix2013,Narita2016} and generates an electric polarization that can be enhanced or reversed by an applied magnetic field \cite{Lee2014,Chaix2016,Ramakrishnan2017}. 

In this Letter, we report the emergence of a novel mesoscopic-scale texture of the 120$^\circ$ helical order in Ba$_3$TaFe$_3$Si$_2$O$_{14}$ (BTFS). This texture is induced by the application of a magnetic field ${\bf H}$ in the $ab$ plane, and is signaled by an extra contribution to the electric polarization measured for $\mu_0H \geq 4$~T. Combining resonant X-ray diffraction (RXD) and neutron scattering experiments, we unveil the magnetic transition behind this singular multiferroic behavior. In accord with the Lifshitz invariants allowed by the chiral $P321$ symmetry, the transition can be explained naturally as a continuous onset of a long-wavelength transverse conical modulation at high field out of the initial 120$^\circ$ helical phase at low fields (double spiral). This mechanism represents a genuine extension of previous models that were restricted to simple antiferromagnets \cite{bogdanov1989} to the more general case of non-collinear spin orders and spiral magnets.  

\begin{figure}[t]
\includegraphics[width=0.5\textwidth]{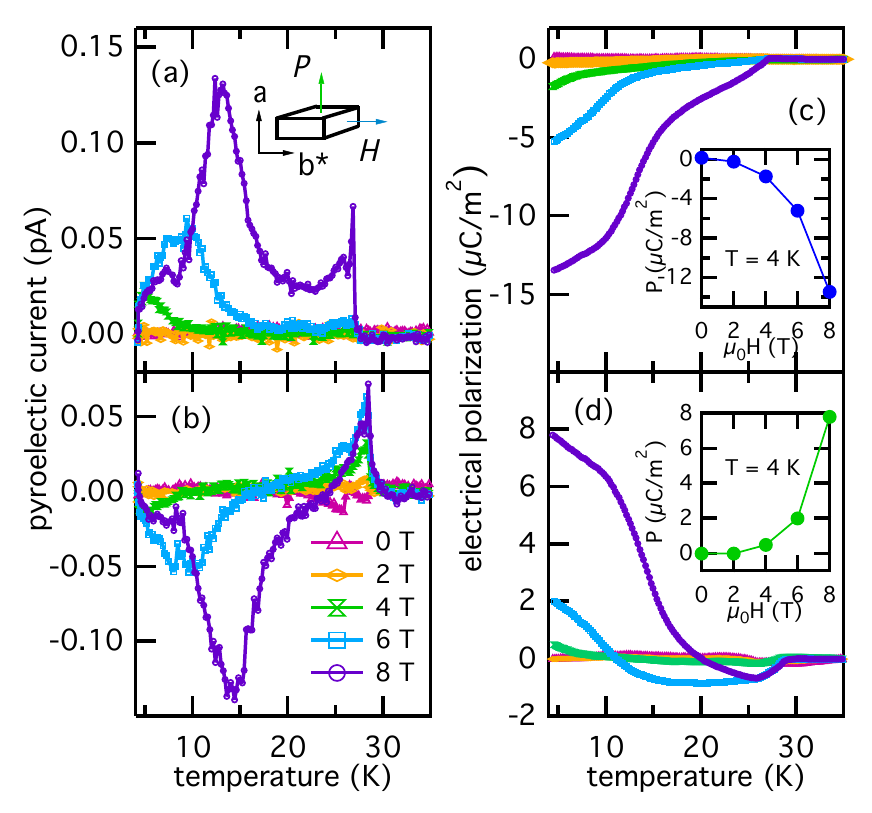}
\caption{
(a-b) Temperature dependence of the pyroelectric current measured along the $a$-axis for several values of the magnetic field applied along $\bf{b}^*$. (a) and (b) correspond to two samples with opposite structural chiralities. The samples were cooled under a small poling field of $\sim$10 kV/m to ensure the selection of a single domain and to suppress a parasitic zero-magnetic--field signal that grows with the poling field strength.
(c-d) Electric polarization obtained by integrating the pyroelectric data in (a-b), respectively. The insets show the electric polarization as a function of the magnetic field at $T =4$~K.} 
\label{fig:poln0}
\end{figure}

Figs. \ref{fig:poln0}(a) and (b) show the pyroelectric current along the $a$-axis as a function of temperature for different values of the magnetic field applied along $\mathbf b^*$ and for the two structural chiralities (see \cite{supp} for experimental details). The measured current develops two clear features that are enhanced by the magnetic field. The first one signals the N\'eel temperature $T_N=28$~K while the second peak appears at 4 T with a maximum at $\sim$ 5 K which shifts to $\sim$ 14 K at 8 T. Interestingly, the relative sign of these features depends on the structural chirality, as can be seen from the comparison between Figs. \ref{fig:poln0}(a) and (b). The corresponding electric polarization is shown in Fig. \ref{fig:poln0}(c) and (d). When the two pyroelectric-current features have the same sign, the electric polarization displays a monotonic behavior as a function of the temperature [Figs. \ref{fig:poln0} (a) and (c)]. However, when the signs are opposite, the behavior is non-monotonic and the polarization is eventually reversed as in \cite{Lee2014} [Figs. \ref{fig:poln0}(b) and (d)]. We note that the maximum component of the polarization is perpendicular to the applied field in the $ab$ plane, which is also the case for $\bf H||a$ (additional measurement details can be found in \cite{supp}).

To understand the origin of the two  electric polarization signals, we looked for possible field-induced changes in the magnetic structure by neutron diffraction. Experiments were performed on the CEA-CRG D23 two-axis diffractometer at the Institut Laue Langevin with an incident wavelength of 1.2794 \AA. We measured BTFS samples with both structural chiralities at 1.5 K and under a magnetic field up to 12 T applied along the ${\bf a}$ and ${\bf b^*}$ directions. We observed a rise of intensity at the zone center reflections, compatible with a ferromagnetic spin component induced by the field along ${\bf a}$ \cite{supp}. The first-order magnetic satellites (0, $k$, $l\pm\tau$) with $\tau$=0.1385, accounting for the helical modulation along the $c$-axis, remain constant for fields up to $\approx$ 4 T before decreasing on further increasing fields [Fig. \ref{fig:neutron} (a)]. We also followed the evolution of second-order harmonics (0, $k$, $l\pm 2\tau$) of pure structural origin in zero field (ascribed to magnetoelastic effects) [Fig. \ref{fig:neutron} (b)] \cite{Chaix2016}. Several of these reflections slightly rise with increasing field due to the onset of a magnetic contribution induced by the deformation of the helix in the applied field. No change of the propagation vector was observed within the accuracy of the experiment, which also holds for $\mathbf H \parallel {\bf b^*}$ and for both structural chiralities. These results clearly indicate that the magnetic structure changes under the applied field above 4 T. 

To gain additional information about the new magnetic structure, complementary small angle neutron scattering (SANS) measurements with high scattering vector $Q$-resolution were performed at the {SANS-I and SANS-II instruments at SINQ, Paul Scherrer Institut with a wavelength of 4.6 \AA\ (see \cite{supp} for more information). The measurements were performed under a magnetic field up to 10.5 T applied along $\bf{b}^*$. The field and temperature dependence of the rocking curve at the $(0,0,\tau)$ position are shown in Figs. \ref{fig:neutron}(c) and (d) respectively. 
At 1.5 K, the $(0,0,\tau)$ satellite develops a clear shoulder when the applied field is higher than 4 T. This shoulder disappears when the temperature is increased above $~$20 K. This feature corresponds to a modulation of the spin structure perpendicular to the $(0,0,\tau)$ reflection propagating along the direction of the applied field $\bf H||b^*$. Remarkably, the period of such a modulation surpasses $350$ nm at 5 T and decreases by increasing the applied field [see Figure \ref{fig:neutron}(e)]. 

\begin{figure}[t!]
\includegraphics[width=0.5\textwidth]{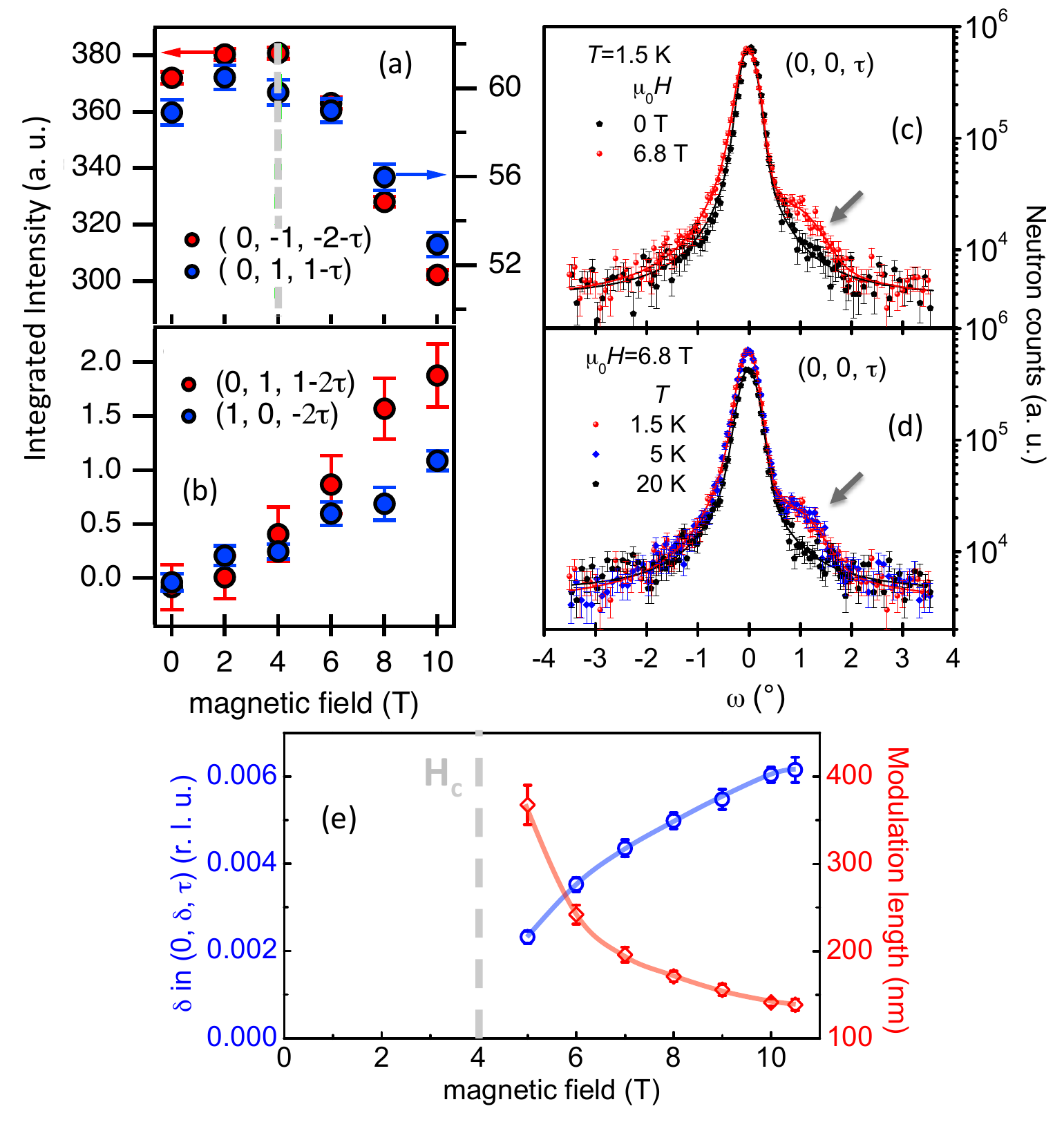}
\caption{(a-b) Field-dependence of the Bragg reflections at T = 1.5 K measured with neutron diffraction on D23 for a magnetic field applied along the $a$-axis. Integrated intensity from rocking curves measured for first order magnetic satellites (a), and for second order satellites (b). (c-d) Rocking curves of the (0, 0, $\tau$) magnetic reflection measured on SANS-I with a magnetic field applied along the ${\bf b^*}$ direction: (c) at 1.5 K under a magnetic field of 0 and 6.8 T; (d) between 20 and 1.8 K under a magnetic field of 6.8 T. The lines are pseudo-Voigt fits. The arrows indicate the presence of the field-induced $\delta$ satellite reflection. {(e) Magnetic field dependence of this long-wavelength modulation wavevector (left axis) and corresponding real space modulation (right axis) measured on SANS-II. The lines are guide for the eyes.}}
\label{fig:neutron} 
\end{figure}

These results are supported by  
RXD experiments at the $L_3$ edge of Fe
using the high-field diffractometer on the UE46-PGM-1 beamline at the synchrotron radiation source BESSY II of the Helmoltz Zentrum Berlin. (See \cite{supp} for detailed information). These experiments reveal that the magnetic moments start to be noticeably affected by a magnetic field larger than 2 T applied in the $ab$ plane. This is reflected by the change in the ratio of scattered intensity of the (0, 0, $\tau$) reflection measured with incident linear horizontal ($\sigma$) and linear vertical ($\pi$) x-ray polarizations, as shown in Figs. \ref{fig:RXD}(a) and (b) for for $\bf H||b^*$ \cite{supp}.  Around 4 T, this fundamental reflection $(0,0,\tau)$ then develops a shoulder, and a further increase in field strength shows a clear satellite that moves away from the main reflection for $\pi$ polarized x-rays [see Fig. \ref{fig:RXD}(a)]. This behavior gives direct 
confirmation for the magnetic phase transition occurring at approximately 4 T. The additional incommensurate component, denoted by $\delta$, emerges along the applied field direction and is labeled $(0,\pm \delta,\tau)$. Note that the slight drift of the main reflection is artificial and caused by small motions of the sample holder. This new magnetic component is perpendicular to the zero field ordering wave vector $\bf \tau$ and reaches $\delta \approx 0.0040 \pm 0.0006$ in reciprocal lattice units at 6.8~T, which corresponds to an incommensurate modulation with real-space periodicity of 240 $\pm$ 40 nm. This value, as well as its dependence on the applied field, is in good agreement with the small-angle neutron scattering experiments [see Fig. \ref{fig:neutron}(e)].

The additional modulation of the zero-field magnetic order is also observed above a magnetic field of 4 T on the $(0, 0, 2\tau)$ reflection [see Figs. \ref{fig:RXD}(c) and (d)], which originates from the magnetic induced  change in the electron density and its associated lattice deformation. Finally it can also be probed on the first order satellite reflection at the oxygen K-edge revealing the sensitivity of the orbital magnetic moment of the oxygen to this new spin texture (see \cite{supp}).

\begin{figure}[t!]
\includegraphics[width=0.5\textwidth]{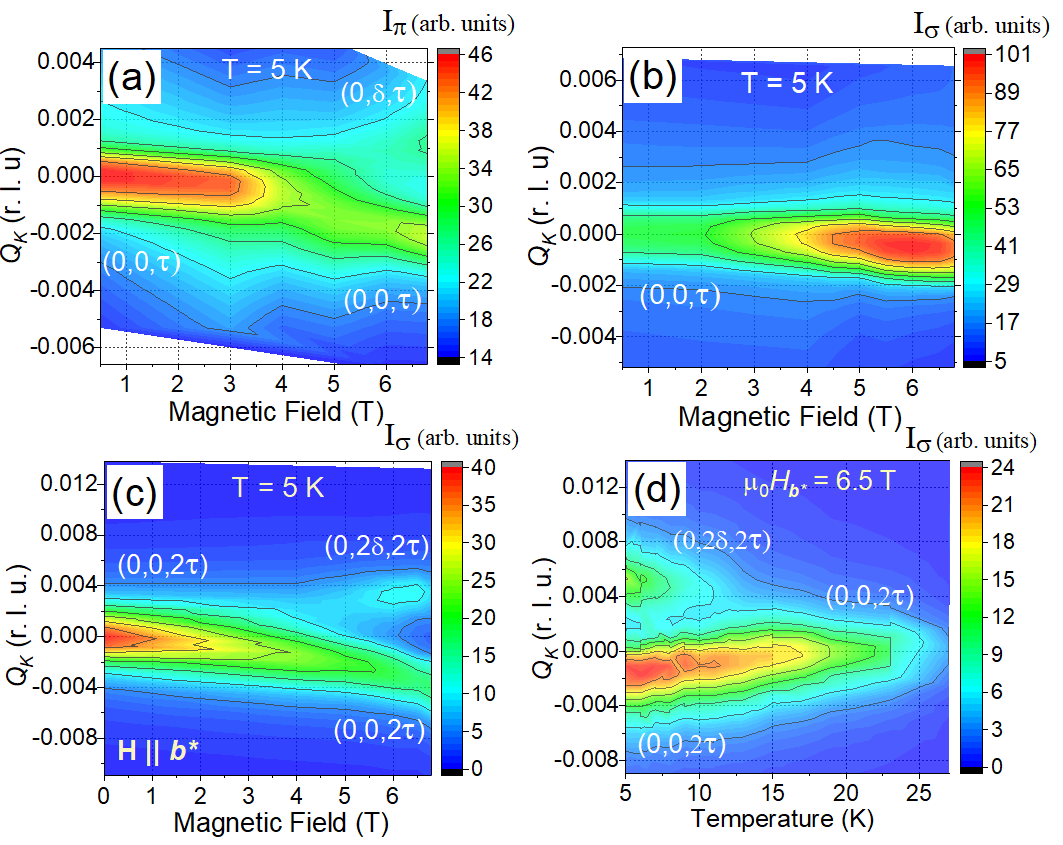}
\caption{{Dependence of the scattered intensity along (0, $Q_K$, $\tau$) as a function of applied field along the ${\bf b^*}$ direction for (a) $\pi$ and (b) $\sigma$-polarized incident x-rays, collected at T = 5 K and E = 709.2 eV. (c) Scattered intensity along (0, $Q_K$, $2\tau$) for $\sigma$-polarized x-rays at T=5 K for $\bf H\parallel b^*$. (d) Temperature dependence of the (0, 0, $2\tau$) reflection in a magnetic field of 6.8 T along the ${\bf b^*}$ direction.}}
\label{fig:RXD}
\end{figure}

Note that the sensitivity of neutron scattering and RXD to the first-order magnetic reflection $(0,0,\tau)$ in zero-field is 
due to different deviations of the magnetic structure from the pure helical state described in Fig. \ref{fig1} (see \cite{supp}) \cite{Scagnoli2013,Chaix2016,Ramakrishnan2017}. Both techniques however probe the evolution of this reflection with magnetic field and converge to show that the magnetic structure is strongly modified in relation to the emergence of a long-wavelength modulation. 

Next, we discuss the origin of this unprecedented magnetic texture. This can be done based on the parametrization introduced in \cite{Reim2018} (see also \cite{dombre1989}), which uses the fact that for the 120$^\circ$ spin order within the Fe triangles $\mathbf S_1 + \mathbf S_2 + \mathbf S_3 = 0 $ and $\mathbf S_1^2 = \mathbf S_2^2 = \mathbf S_3 ^2$: 
\begin{align}
\begin{pmatrix}
\mathbf S_1\\
\mathbf S_2\\
\mathbf S_2\\
\end{pmatrix}=
\begin{pmatrix}
\mathbf V_1\\
 - {1\over 2} \mathbf V_1 + {\sqrt{3}\over 2}\mathbf V_2\\
-{1\over 2} \mathbf V_1 - {\sqrt{3}\over 2}\mathbf V_2\\
\end{pmatrix},
\end{align}
where the vectors $\mathbf V_1 = (X_1,Y_1,Z_1)$ and $\mathbf V_2 = (X_2,Y_2,Z_2)$ are such that $\mathbf V_1^2 = \mathbf V_2^2$ and $\mathbf V_1 \cdot \mathbf V_2 = 0$. In our system, the frustrated exchange interactions lead to the 120$^\circ$ order that displays a helical modulation along the $c$-axis with the wavevector $\tau = 0.1385$. In addition, the spins tend to lie in the $ab$ plane due to the Dzyaloshinskii-Moriya interactions within the Fe triangles \cite{Chaix2016}. This zero-field helical spiral is described by  $\mathbf{V}_1 = \hat{\mathbf{x}} \cos \phi + \hat{\mathbf{y}} \sin \phi$ and $\mathbf{V}_2 = - \hat{\mathbf{x}} \sin \phi + \hat{\mathbf{y}} \cos \phi$ with $\phi = \tau z$, where the $y$-axis corresponds to the two-fold symmetry axis along $\hat {\mathbf a} + \hat{\mathbf b}$, $\hat {\mathbf x}$ is perpendicular to the $y$-axis in the $ab$-plane, and $z$ is along the $c$-axis (see Fig.~\ref{fig1}). Above a critical field $ {\bf H} \| {\bf b^\ast}$, the helical spiral is expected to transform into a cycloidal one \cite{Mostovoy2006} with $\mathbf{V}_1 = \hat{\mathbf{z}} \cos \phi + \hat{\mathbf{a}} \sin \phi$ and   $\mathbf{V}_2 = - \hat{\mathbf{z}} \sin \phi + \hat{\mathbf{a}} \cos \phi$. 

The lack of inversion symmetry of the langasite crystal lattice allows for an additional modulation of the spiral state, which we assume to be the rotation around an axis in the $ab$-plane, $\mathbf{n} = (\cos \chi, \sin \chi, 0)$,  through an angle $\psi$.  The Lifshitz invariants favoring this type of modulation with an in-plane propagation vector are given in \cite{supp}. Microscopically, these invariants trace back to Dzyaloshinskii-Moriya couplings in the $ab$ plane between Fe spins of different triangles not included in previous models \cite{Marty2008,Loire2011,Zorko2011,Scagnoli2013,Marty2010}. For the cycloidal spiral, the first three of them vanish after averaging over $\phi$ that varies on a much smaller length scale than $\psi$, as the  $\phi$-rotation originates from frustrated exchange interactions. The fourth and fifth invariants are proportional to $ 
[1+\sin^2( \chi -\tfrac{\pi }{6})](\mathbf{n} \cdot \mathbf{\nabla}) \psi$ 
and 
$  \sin (\chi-\frac{\pi }{6} )(\cos 2 \chi \partial_x \psi - \sin 2 \chi \partial_y \psi)$ respectively.
Both these interactions favor the additional helical modulation of the spin order with a propagation vector parallel to $\mathbf{n} = \hat {\mathbf b}^{*}$, and hence are likely behind the field-induced spin texture observed in our system. A sketch of such a new magnetic phase is shown in the inset of Fig.~\ref{fig:phdiag} (see also Supp. Mater. \cite{supp}).

The energy decrease due to the additional rotation must overcome an increase of the anisotropy energy, which is why this rotation appears at the spiral flop transition where the anisotropy  gap is reduced by the magnetic field. This scenario represents a generalization to the case of non-collinear spiral magnets of the magnetic-field-induced modulations discussed in \cite{bogdanov1989} for simple antiferromagnets. The ingredients  required for the additional modulation with an in-plane propagation vector set the stage for the emergence of more complex objects such as skyrmions and merons \cite{bogdanov1998,bogdanov2002,Rybakov2015,kharkov17}. 
 
The intriguing behavior of the field-induced electric polarization, $\bf P$, may now be understood in terms of two different mechanisms related to the magnetic structure. The first one emerges right below $T_\mathrm{N}$ and is connected to the field-induced deviation from the 120$^\circ$ magnetic arrangement as proposed in Ref. \onlinecite{Chaix2016}. This mechanism, however, is not expected to depend on the chirality of the crystal structure. The chirality-dependent in-plane polarization induced by the second modulation can be qualitatively understood as a linear magnetoelectric effect. We note that, even if the initial order below the threshold magnetic field preserves all the symmetry elements of the $P321$ space group, the magnetoelectric coupling $-g_1(E_x H_x + E_y H_y)$ is allowed, which results in ${\bf P} \| {\bf H}$. 
In its turn, the additional modulation above the critical magnetic field  breaks the two-fold axis $2_y$, which then allows for a second magnetoelectric coupling $-g_2 (E_x H_y - E_y H_x)$, giving rise to the electric polarization ${\bf P} \perp  {\bf H}$, as observed in the experiments.  
Symmetry analysis of multiferroic interactions (see \cite{supp}) also  gives the electric polarization ${\bf P} \perp  {\bf H}$. This magnetically-induced electric polarization appears only because of the lack of inversion symmetry in the langasite crystal lattice. Therefore, the sign of $P$ is determined by the sign of the crystal chirality. The occurrence of polarization caused by a mesoscopic spin texture in chiral structures add to other novel mechanisms in chiral magnetoelectrics such as MnSb$_2$O$_6$ \cite{Johnson2013, Kinoshita2016}, where the sign of the spin spiral tilting defines the electric polarization. 

\begin{figure}[t!]
\includegraphics[width=0.4\textwidth]{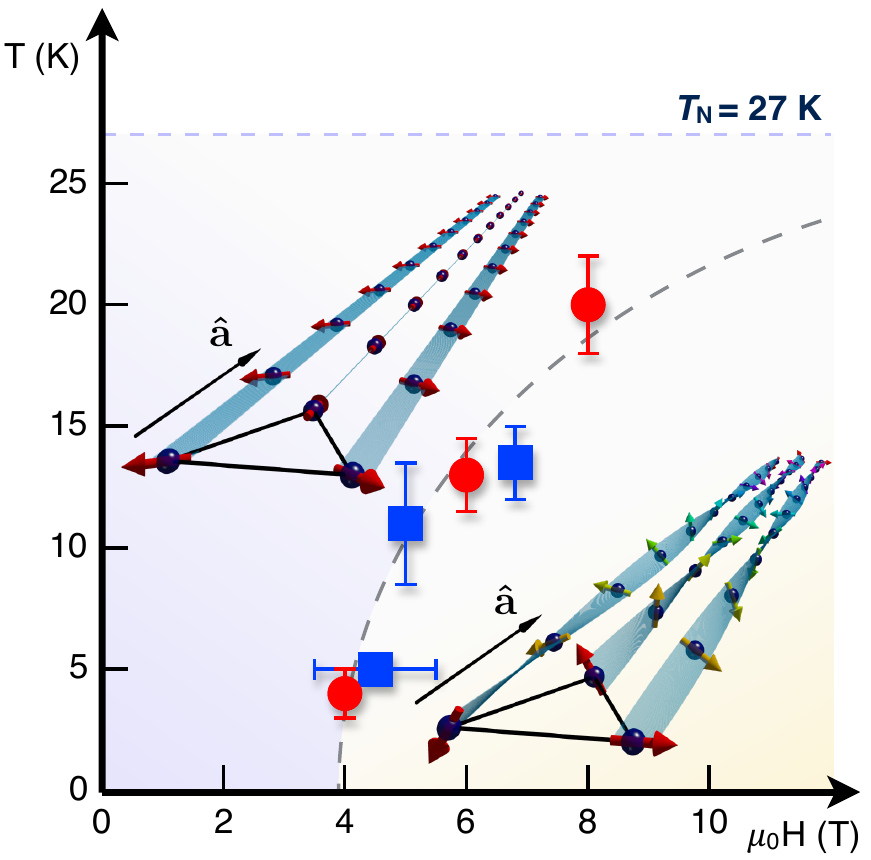}
\caption{$H$-$T$ phase diagram of BTFS showing the field-induced mesoscopic spin texture phase. The blue squares and the red circles represent the transition points determined by RXD and pyroelectric current, respectively. They also fully agree with the SANS measurements. The insets show an oversimplified picture of the proposed zero-field and field-induced magnetic arrangements in the $ab$ plane. Note that this is drawn for a field applied along $\bf a$ for clarity purposes but the results are fully similar for $\bf H||b^*$. Both structures are detailed in the supplementary information \cite{supp}. }
\label{fig:phdiag}
\end{figure}

To confirm that the new spin texture observed in the BTFS langasite is indeed behind the polarization features observed in Fig. \ref{fig:poln0}, the appearance of the extra components on the $(0,0,\tau)$ and $(0,0,2\tau)$ reflections observed by RXD was mapped out throughout the $(H,T)$ phase diagram of BTFS in Fig. \ref{fig:phdiag}. As we see, the onset field of the $\delta$-modulation is found to perfectly match the occurrence of the second component of the pyroelectric current and relates to the inflection points in the net polarization. We note that this transition can also be identified by an anomaly of the elastic constants \cite{Quirion2017}.

In summary, we report 
a magnetic field-induced phase transition in Ba$_3$TaFe$_3$Si$_2$O$_{14}$ resulting in the secondary long-wavelength modulation of its spin structure, which is new and fundamentally different from those found in other chiral magnets. It can be understood as superstructural spiral modulation that is perpendicular to the basic short-period spiral modulation of the 120$^\circ$ basal spin ordering and can be described as a ``multi-q" structure with two remarkably different wavevectors. At this transition we also find an emergent chirality-dependent component of electric polarization induced by the additional magnetic modulation.  Using a phenomenological approach, we explain the transition to the phase with a mesoscopic spin-texture propagating in the $ab$ plane that interestingly contains all the necessary ingredients for the formation of antiferromagnetic skyrmions. We anticipate that novel topological objects and exotic textures can be realized in langasites and other non-centrosymmetric frustrated magnets with similar symmetry properties. In this respect, it will be interesting to study topological properties of this new state. 

\paragraph{Acknowledgements.---}
M.R. and E.C. contributed equally to this work. We thank P.M. Derlet for insightful discussions, K. Rolfs and E. Pomjakushina for their assistance in sample characterization. We acknowledge A. Hadj-Azzem and J. Balay for their help in the preparation of the samples and A. A. Mukhin for fruitful discussions. MR, NG and JW acknowledge financial support of the Swiss National Science Foundation (SNSF) (Sinergia project 'Toroidal moments' No. CRSII2\_147606),  (project No. 200021\_162863) and Sinergia project 'NanoSkyrmionics' (grant CRSII5\_171003), respectively. This project has received funding from the European Union Horizon 2020 research and innovation program under grant agreement No 730872 CALIPSOplus. It was also financially supported by Grant No. ANR-13-BS04-0013. E.C. has benefited from a PRESTIGE (No. 2014-1-0020) fellowship and CMIRA-Accueil Pro-2015 funding for this research work. 

\bibliography{BTFS_PRL}

\end{document}